\documentclass[twocolumn,showpacs,preprintnumbers,amsmath,amssymb]{revtex4}
\usepackage{amssymb}
\usepackage[dvips]{graphicx}
\begin{document}

\title{Vortex matter in the charged Bose liquid at absolute zero}
\author{V. V. Kabanov$^1$ and A. S. Alexandrov$^2$ }

\affiliation{$^{1}$Josef Stefan Institute 1001, Ljubljana, Slovenia \\
$^2$Department of Physics, Loughborough University, Loughborough, United Kingdom}

\begin{abstract}
The Gross-Pitaevskii-type equation is solved for the charge Bose liquid in an
external magnetic field at zero temperature. There is a vortex lattice
 with  locally broken charge neutrality.  Remarkably, there is no upper  critical field at zero
temperature,  so  the density of single flux-quantum vortices monotonously increases with
the magnetic field up to $B=\infty$ and no indication of a phase transition. The size of
each vortex core decreases as about $B^{-1/2}$ keeping the system globally charge neutral. If bosons are composed of two fermions,  a phase transition to a spin-polarized Fermi liquid at some magnetic field larger than the pair-breaking field is predicted.
\end{abstract}


\maketitle

Charged Bose liquids (CBLs) have been solely of academic interest for a long time
\cite{sha,fol,fet,bis,res,lee,bru,woo}. Notwithstanding, experimental realization of the
Bose-Einstein condensation (BEC) of trapped ultra-cold atoms 
\cite{and,dav,bra,mew,hag,exp} made it possible to create ultracold plasmas \cite{kil} by using lasers to trap and cool neutral atoms to temperatures of 1 mK or lower. Another laser then ionizes the atoms by giving each of the outermost electrons just enough energy to escape the electrical attraction of its parent ion.
 The ions retain the millikelvin temperatures of the neutral atoms, so  they may  bose-condense, if their spin is an integer.
There is also  growing experimental evidence for preformed 2e-charged bosons in
 high-temperature cuprate superconductors, such as  normal state pseudogaps, unusual upper critical fields,   small normal-state Lorentz numbers, etc  \cite{alebook}.    Similar charged boson physics is expected in a lattice of mesoscale superconducting dots, if parameters are chosen such that single-electron tunnelling is suppressed and only 
Cooper-pair charges tunnel  between the domains via Josephson tunnelling \cite{shon}.
It is also possible to describe the universal features of the superconductor-insulator transition as a function of disorder in quasi-two dimensional systems in terms of  boson physics \cite{fis,wal}. In order to model the transition in terms of
bosons, one has to include the Coulomb repulsion, otherwise all
bosons would collapse into the lowest lying highly localised state.  

These developments have renewed interest in CBL as a fundamental reference system. A \emph{non-interacting} gas of charged bosons cannot bose-condense at any
 finite magnetic field because of a one-dimensional character of motion in the lowest
 Landau band \cite{sha}. However, \emph{interacting} charged bosons  condense below some
(upper) critical field $B\leqslant H_{c2}(T)$ since their
collisions remove the one-dimensional singularity of  the density of
states \cite{ale}. The BEC field diverges with decreasing
temperature \cite{ale,alekabbil}, so that $H_{c2}(T)=\infty$ at
absolute zero. A single vortex in CBL has  a charged core and an
electric field inside \cite{alevor}, while its magnetic field is
virtually identical to  the Abrikosov vortex \cite{abr}.

 Here we present the ground state of CBL in an arbitrary magnetic field  solving
numerically the Gross-Pitaevskii  -type equations with the long-range Coulomb
interaction between bosons. We find a lattice of
charged vortices, which does not disappear at any finite magnetic
field. The density of  vortices monotonously increases and their
core size decreases with the magnetic field up to $B=\infty$. The size of vortices also  depends on the
thickness of CBL films different from the conventional superconducting films. When bosons are composed of two fermions, there is  a phase transition to a spin-polarized Fermi liquid at some magnetic field.

 The  Hamiltonian  of charged bosons  on
a compensating homogeneous background (to ensure global charge neutrality) in the
external magnetic
field  with the vector potential ${\bf A}({\bf r})$ is given by

\begin{eqnarray}
H&=&\int d{\bf r} \psi^\dagger({\bf r})\left[-{(\hbar {\bf \nabla}-i e
{\bf A}/c)^2
\over{2m}}
-\mu \right]\psi({\bf r})\cr
&+&
\frac 1 2 \int d{\bf r} \int d{\bf r'} V({\bf r} - {\bf r'})\cr
&\times&[\psi^\dagger({\bf r})
\psi^\dagger({\bf r'})\psi({\bf r'})\psi({\bf r})-2n\psi^\dagger({\bf r})
\psi({\bf r})],
\end{eqnarray}
where $m,e,n,\mu$ are the mass, charge, average density and  chemical
potential of bosons, respectively, and $V({\bf r})=e^2/|{\bf r}|$ is
their Coulomb repulsion \cite{ref}.

The equation of motion for the Heisenberg field operator,
$\psi({\bf r},t)$, is derived using this Hamiltonian. If the
density is relatively high, so that the dimensionless Coulomb
repulsion $r_{s}=me^{2}/\hbar^{2}(4\pi n/3)^{1/3}$ is not large,
 one can expect that the occupation
 numbers of
 one-particle states are not very much different from those in
the ideal Bose-gas. In particular, one state remains 
macroscopically occupied at $T=0$K. Then, following Bogoliubov
\cite{bog} one separates the large matrix element $\psi_{0}$ from
$\psi$ by treating the rest $\tilde{\psi}$ as small fluctuations,
$\psi({\bf r},t)=\psi_{0}({\bf r})+\tilde{\psi}({\bf r},t)$.
The anomalous average $\psi_{0}({\bf r})=\langle \psi({\bf
r},t)\rangle$ is approximately equal to $\sqrt{n}$ in a
homogeneous system at $T=0$. Substituting the Bogoliubov
displacement transformation into the equation of motion
and collecting $c-number$ terms of $\psi_0$, one obtains  the
 equation for the macroscopic condensate
wave function as \cite{alevor}
\begin{eqnarray}
&&\left[{(\hbar {\bf \nabla}-i e {\bf A}/c)^2\over{2m}} +
\mu\right]\psi_0({\bf r})\cr
&=&\int d{\bf r'} V({\bf r} - {\bf r'})[\psi_0^{*}
({\bf r'})\psi_0
({\bf r'})-n]\psi_0({\bf r}).
\end{eqnarray}

The integro-differential  equation (2) is quite different from
original Ginzburg-Landau  (GL)\cite{gin} and the Gross-Pitaevskii \cite{gro}
equations, describing the order parameter in the BCS and neutral
superfluids, respectively. As recognised by  one of us \cite{alevor}
 the coherence
 length in CBL is just the same as the screening radius, so the core of a single vortex is charged \cite{ref2}.
Indeed, introducing  the dimensionless quantities: $f=|\psi_{0}|/n^{1/2} $,
${\bf \rho}={\bf r}/\lambda $,
${\bf h}=e\xi\lambda curl{\bf  A}/\hbar c$ for the order
parameter, length and magnetic field, respectively, one obtains the coherence length
 about the same as the  screening radius at $T=0$K
\cite{res}, $\xi=(\hbar/2^{1/2}m^{*}\omega_{p})^{1/2}$, where
$\omega_{p}=(4\pi n e^{2}/m)^{1/2}$ is the zero-temperature
plasma frequency \cite{fol}. The London penetration
depth is conventional, $\lambda= (mc^{2}/4\pi ne^{2})^{1/2}$, but a new feature is an
 electric field potential, $e\phi({\bf r})= \int d{\bf r'} V({\bf r} - {\bf r'}) [|\psi_0
({\bf r'})|^{2}-n].$ Moreover, the chemical
potential $\mu$ is zero, as it should be in the globally neutral CBL
in the thermal equilibrium below the BEC critical temperature.
\begin{figure}
\begin{center}
\includegraphics[angle=-0,width=0.47\textwidth]{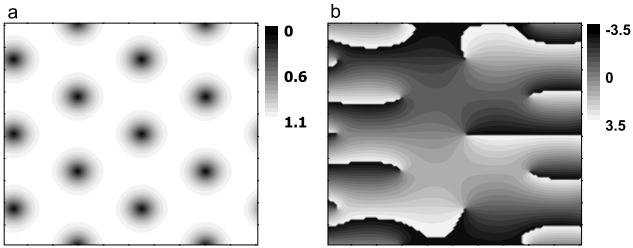}
\vskip -0.5mm \caption{A few vortices in a sample of the
size $L=22.4 \xi$, (a). The phase profile of the order parameter is
shown in (b). The phase changes by $2\pi$ around each core at any
magnetic field. }
\end{center}
\end{figure}

Any realistic CBL is an extreme type II  with a very large
Ginzburg-Landau parameter, $\kappa=\lambda/\xi \ggg 1$
\cite{alevor}. For example, the coherence length and the electric field inside
the vortex core are about 1 nm or less and 10 mV, respectively,
with the material parameters typical for cuprates ($m=10 m_{e}$,
$n=10^{21}cm^{-3}$ and $\epsilon_{0} \gtrsim 100$),
and $\kappa $ is about $10^3$ with these
parameters. Hence, the   magnetic field is practically
homogeneous, and the ground state $\psi_0({\bf r})$ can be found
 by minimizing the energy functional $E$ with respect
to $\psi_0({\bf r})$,
\begin{eqnarray}
&&E(\psi_0)={1\over{2m}}\int d{\bf r} |(\hbar {\bf \nabla}-i e
{\bf A}/c)\psi_0({\bf r}|^2 \cr
&+&
\frac 1 2 \int d{\bf r} \int d{\bf r'} V({\bf r} - {\bf r'})
|\psi_0({\bf r})|^2(|\psi_0({\bf r'})|^2-2n),
\end{eqnarray}
where ${\bf A}=\{0,Bx,0 \}$. In numerical simulations we consider
a sample with the rectangular cross-section  $L\times L$ and the magnetic
flux $BL^{2}=m\Phi_{0}$, where $m$ is an integer ($\Phi_{0}$ is
the flux quantum). When the magnetic field ${\bf B}$ is applied along z-direction, the order parameter $\psi_{0}(x,y)$
does not depend on $z$ obeying the following translation symmetry,
\begin{eqnarray}
&&\psi_{0}(x+L,y)= \exp{(-ieBLy/\hbar c)}\psi_{0}(x,y) \cr
&&\psi_{0}(x,y+L)= \psi_{0}(x,y).
\end{eqnarray}
These relations can be  used as  boundary conditions when m is an integer.

Because Eq.(3) does not contain the penetration
depth, it is convenient to  introduce new dimensionless coordinates
${\bf x}={\bf r}/\xi$, the vector potential, ${\bf a}=(0, 2\pi
Bx\xi^{2}/\Phi_{0},0)$, and the Coulomb energy $v({\bf x})= e\phi /(\omega_{p}^{2} m^{*}
\xi^{2})$. As a result, the problem is reduced to minimization of
the functional 
\begin{eqnarray}
&&E(f)=\frac {\hbar^2 n \xi}{2m}\int d{\bf x}
\Bigl [|( \nabla-i{\bf a} )f({\bf x})|^2 \cr &+& v({\bf
x})(|f({\bf x})|^2-1)\Bigr ],
\end{eqnarray}
where the Coulomb field satisfies the Poisson
equation, 
\begin{eqnarray}
\Delta v({\bf x}) &=& 1-|f({\bf x})|^{2}.
\end{eqnarray}
To compare CBL vortex state  with the  Abrikosov vortex lattice we also minimize
the conventional GL  functional using the same dimensionless unites, 
\begin{eqnarray}
&&E_{GL}(f)=\frac {\hbar^2 n_s \xi}{2m} \int d{\bf
x} \Bigl [|( \nabla-i{\bf a} )f({\bf x})|^2 \cr &-& |f({\bf
x})|^{2}+\frac 1 2 |f({\bf x})|^{4}\Bigr ],
\end{eqnarray}
where $\xi=\hbar^2/(2m |\alpha|)^{1/2}$,  $n_s=|\alpha|/\beta$ and the order parameter $f$ is normalised by $\sqrt{n_s}$. Here $\alpha$ and $\beta$ are conventional GL coefficients \cite{gin}. We apply the  standard discretization procedure 
 described in
Ref.\cite{bonca}. Eq.(6) for the electrostatic potential 
is solved by the Fourier transform in the
discrete form, and the resulting
energy is minimized with the conjugated gradient algorithm.

\begin{figure}
\begin{center}
\includegraphics[angle=-0,width=0.47\textwidth]{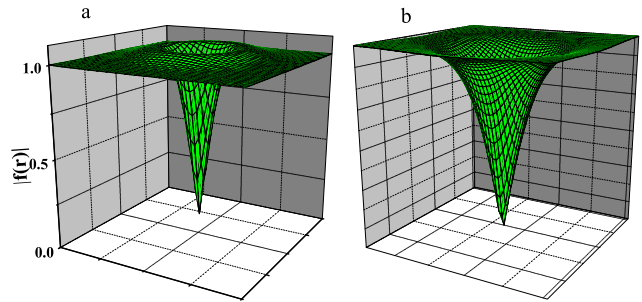}
\vskip -0.5mm
\caption{Single vortex in CBL \cite{alevor} (a) compared with the Abrikosov vortex (b).
}
\end{center}
\end{figure}

Since both functionals depend only on the  dimensionless
vector-potential ${\bf a}$ which is proportional to the product
$B\xi^{2}$,  simulations can be performed at
fixed $L$ and $\xi$ by changing $B$ or  at fixed $L$ and
$B$ by changing $\xi$. Our numerical results are shown in
Figs. 1-4. At any value of the magnetic field we find the
triangular vortex lattice. While the field is small, there are
only a few vortices per sample cross-section, Fig.1a. When vortices
are far apart, their interaction yields only a small contribution
to the energy functional but even in that case a triangular lattice
of vortices is clearly seen in CBL, Fig.1. 

Each vortex carries one flux quantum, 
as can be seen
from the phase profile in Fig.1b.   It has an unusual core, Fig.2,
in agreement with Ref. \cite{alevor}, which differs qualitatively
from the Abrikosov vortex \cite{abr} due to a local charge
redistribution caused by the magnetic field. 
\begin{figure}
\begin{center}
\includegraphics[angle=-0,width=0.47\textwidth]{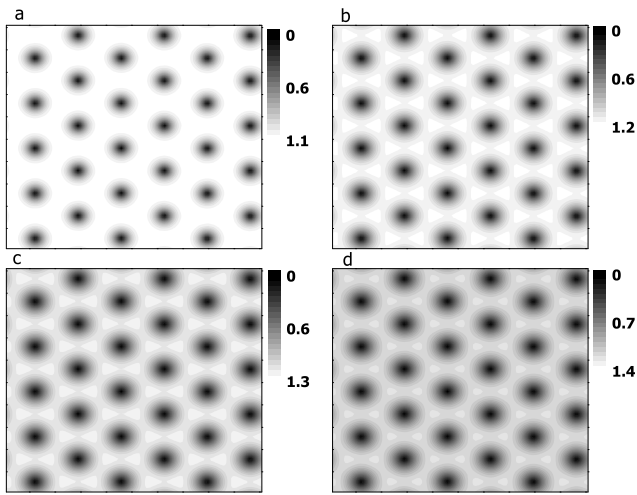}
\vskip -0.5mm \caption{The vortex lattice in CBL for 30 flux
quantum per cross-section, (a) $L/\xi=33.67$, (b) $L/\xi=25.25$, (c)
$L/\xi=14.43$, (d) $L/\xi=10.1$ One can see from the scale near each figure  that the order
parameter remains large outside the cores, $f> 1$  at any $\xi$ (or at any magnetic field).}
\end{center}
\end{figure}
The
breakdown of the local charge neutrality, Fig.2, is due to the absence of an
 equilibrium $normal$ state solution in CBL  at $T=0$ with $\psi_0=0$, as explained in Ref.\cite{alevor}.
  
\begin{figure}
\begin{center}
\includegraphics[angle=-0,width=0.47\textwidth]{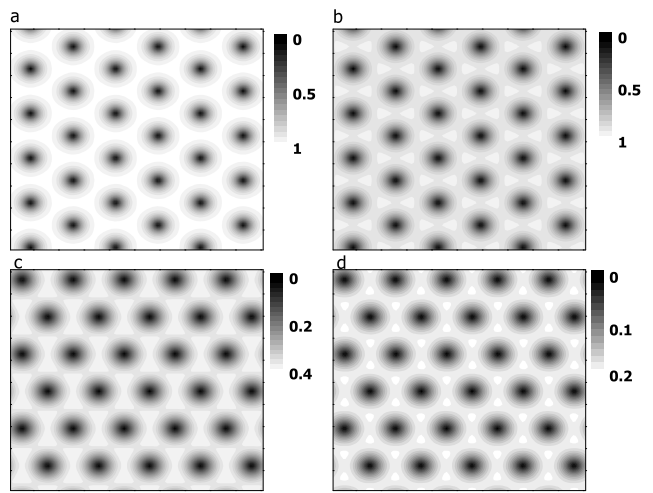}
\vskip -0.5mm \caption{The Abrikosov vortex lattice  for 30 flux
quantum per cross-section (a) $L/\xi=33.67$, (b) $L/\xi=25.25$, (c)
$L/\xi=14.43$, and for (c) $L/\xi=13.87$, which corresponds to $B$ close to $H_{c2}=\Phi_0/(2\pi \xi^2)$. The order parameter decreases when $B$ approaches the conventional upper critical field.}
\end{center}
\end{figure}

 Increasing the
 field  first increases the vortex
density with about constant size of the cores, as in  conventional
superconductors, Fig.3 and Fig.4. However, quite different from the Abrikosov lattice, 
increasing  the field further does not lead to a superfluid
to normal phase transition, but instead it increases the density
of vortices by decreasing the size of every individual core,
Fig.3c,d. Remarkably, each vortex carries one flux quantum at any field.  Keeping the global
charge neutrality the charge heterogeneity depends on the magnetic
field, and the core diameters decrease with the field, when the
field is large, $\xi^2>2\pi \hbar c/(eB)$. Indeed, in this regime
the "bare" coherence length $\xi$ becomes irrelevant, but the only
characteristic length is  the
distance between single flux-quantum vortices, i.e. $r \approx
\sqrt{2\pi \hbar c/(eB)}$. As a result, the amplitude real-space
modulations of the order parameter increase with the magnetic field
in CBL, while they decrease in conventional superconductors, where the
 order parameter vanishes at and above the finite $H_{c2}=\Phi_0/(2\pi \xi^2)$
(fig.4c,d). 

There is another difference between CBL and conventional vortex matter in  case of
thin films. If we assume that the film
thickness $d$ is small, $d\ll\xi$, then the left hand side of Eq.(6) takes the form
 $(1-|f({\bf x})|^{2})d\delta(z)/\xi$.  The dimension analysis readily shows
that the true coherence length, $\xi_{2D}$  depends on the thickness as
$\xi_{2D}=(\xi^{4}/d)^{1/3}$ in that case.  As a result the size of vortex cores  depends on the
thickness of CBL films different from the conventional films.

There is also  an important consequence of the infinite (orbital) upper critical field at absolute zero in such CBLs, where singlet bosons
are formed of two fermions \cite{alebook}. In this case sufficiently large magnetic field can break  bound pairs via a spin-flip of one of two fermions, if  triplets are unstable, because the singlet binding energy $\Delta$ decreases with the field as $\Delta(B)=\Delta-2\mu_B B$ ($\mu_B=e\hbar/(2m_e)$ is the electron Bohr magneton) \cite{zav}.  A spin-polarised Fermi liquid appears at $B\geqslant H_p$, where $H_p=\Delta/(2 \mu_B)$ is the pair-breaking field. In this estimate we neglect the orbital (Landau) diamagnetism of bosons and fermions, and the Coulomb energy of 
the charged-modulated vortex lattice. The latter is of the order of  $e\phi_c n\xi^2 B/\Phi_0$ per unit volume, where $\phi_c\sim \hbar^2/(e m\xi^2)$ is the characteristic  electrostatic potential inside  vortex cores. The Coulomb energy is small compared with the spin (Pauli) contribution if $m_e/m \ll 1$,  which we assume to be the case, so    diamagnetic contributions are also small. However,  bound pairs  still survive up to a higher field $H^*=H_p+n/(N\mu_B) > H_p$ due to the Pauli exclusion principle, which prevents any further decay of pairs, if the number of fermions $\thickapprox N|\Delta(B)|$ remains smaller than $2n$ ( $N$ is the fermion density of states). There is a boson-fermion mixture, if $H_p < B < H^*$, with the fermion density  modulated in real space because of charged vortices. Normal fermions (as well as normal bosons pushed up from the condensate by temperature)  are 
distributed inhomogeneously  across the sample with the maximum 
density in the vortex cores, where their potential energy  is at 
minimum. The excess density of normal carriers inside the cores  screens the electric field caused by the inhomogeneous condensate density. If the screening length due to  normal fermions becomes smaller than the coherence length $\xi$, one can  expect a nontrivial field dependence of the  size of vortices, which disappear at $B=H^*$.

In conclusion, we have   found the
triangular  lattice of single-flux-quantum charged vortices in CBL which
 cannot be destroyed by any magnetic field at zero
temperature. The vortex density  monotonously increases and their
core size decreases with the magnetic field up to $B=\infty$ with
no indication of a phase transition. The core size depends on  the
thickness of CBL films.  At finite temperatures 
 $H_{c2}(T)$ is finite \cite{ale,alekabbil}. Nevertheless,
unusually large charge modulations with the scale depending on  the
magnetic field should persist at finite temperatures as well.  The phase transition to the spin-polarized Fermi liquid at some magnetic field larger than the pair-breaking field has been predicted for preformed bosonic pairs. These results are relevant for  real charged Bose-liguids in ultracold plasmas and in the superconducting cuprates, and for  superconducting quantum dots and  
superconductor-insulator phase transitions described by a similar boson physics. There is also a close analogy between the vortex structure in CBL and the
Josephson vortices. Since the normal phase is not defined  below $T_c$, there is no
"normal" vortex core in CBL, and there is no "normal" core  in the Josephson vortex either.  One can
define the lower critical field $H_{c1}$ when a first vortex penetrates into  CBL \cite{alevor} and into the Josephson junction \cite{schmidt}, but the upper critical field is infinite in both cases.


\begin{thebibliography}{90}
\bibitem{ref}  The Coulomb repulsion of preformed pairs is $4e^2/r\epsilon_0 $,
where $\epsilon_0$ is the static dielectric constant of the host material
and $e$ is the elementary charge.
\end{thebibliography}
\end{document}